\documentstyle[11pt,aaspp4,epsf]{article}
\def\etal   {{\rm ~et al.,~}}
\def\kms    {\ifmmode{{\rm ~km~s}^{-1}}\else{~km~s$^{-1}$}\fi}
\def\lsun   {\ifmmode{{\rm ~L}_\odot}\else{~L$_\odot$}\fi}
\def\msun   {\ifmmode{{\rm ~M}_\odot}\else{~L$_\odot$}\fi}
\lefthead{Greenhill\etal}
\righthead{Water Maser Emission in NGC~1068}

\begin{document}

\title{VLBI Imaging of Water Maser Emission from the Nuclear Torus of NGC~1068}

\author{L. J. Greenhill}
\affil{Harvard-Smithsonian Center for Astrophysics, 60 Garden St, 
Cambridge, MA 02138}

\author{C. R. Gwinn, R. Antonucci}
\affil{Department of Physics, Broida Hall, University of California, 
Santa Barbara, CA 93106}

\and
\author{R. Barvainis}
\affil{Haystack Observatory, NEROC, Route 40, Westford, MA 01886}

\begin{abstract}

We have made the first VLBI synthesis images of the H$_2$O maser emission
associated with the central engine of the Seyfert galaxy NGC~1068. Emission
extends about $\pm300$\kms~from the systemic velocity. Images with
submilliarcsecond angular resolution show that the red-shifted emission lies
along an arc to the northwest of the systemic emission. (The blue-shifted
emission has not yet been imaged with VLBI.) Based on the maser velocities and
the relative orientation of the known radio jet, we propose that the maser
emission arises on the surface of a nearly edge-on torus, where physical
conditions are conducive to maser action. The visible
part of the torus is axially thick, with comparable height and radius. The
velocity field indicates sub-Keplerian differential rotation around a central mass of $\sim 1
\times 10^7$ M$_\odot$ that lies within a cylindrical radius of about 0.65 pc.
The estimated luminosity of the central engine is about 0.5 of the
Eddington limit. There is no detectable compact radio continuum emission near 
the proposed center of the torus ($T_B< 5\times10^6$ K
on size scales of $\sim 0.1$ pc), so that the observed flat-spectrum
core cannot be direct self-absorbed synchrotron radiation.

\end{abstract}


\section{Introduction}

The galaxy NGC~1068 is widely believed to harbor a Seyfert I nucleus that is
obscured by a dusty edge-on torus (\cite{Ant93}, and references therein). 
X-rays of energies up to 10 keV from the central engine of 
the active galactic nucleus (AGN)
are blocked by an atomic column density of at least $10^{25}$ cm$^{-2}$
(\cite{Mul92}). The inferred axis of the torus is nearly north-south, and 
broadline optical and
ultraviolet emission is scattered into the line of sight on angular
scales as small as $0\rlap{.}''1$ (\cite{AM85}, \cite{AHM94}, \cite{Cap95}). 
An ionization
cone is visible in [OIII] light (\cite{Evans91}, \cite{Mac94}) and is closely
associated with a roughly north-south radio jet (\cite{UNW87}, \cite{Mux96}). 
A jet component near the southern end, at the
base of the cone has a rising spectrum, suggesting that the central engine
lies approximately there (\cite{Galcont}), 
as ultraviolet polarization measurements also indicate
(\cite{AHM94}, \cite{Cap95}).

A water maser source (\cite{CHL84}) lies at the suggested
location of the central engine (\cite{CL86}, \cite{Galh2o}). 
Claussen \& Lo (1986) showed that the H$_2$O maser is compact on parsec scales
and first suggested that the dusty molecular torus of the AGN 
is responsible for the maser emission.  
Thus, the structure and velocity field of the 
torus might be traced by the maser, which can be studied with Very Long Baseline
Interferometry (VLBI) on scales as small as 0.1 milliarcseconds (mas), despite
high foreground column densities. At a distance of 15 Mpc, 
0.007 pc subtends 0.1 mas.
The H$_2$O maser emission marks the presence of warm ($\sim 400$ K), high
density ($n_{{\rm H}_2}\sim 10^8 - 10^{10}$ cm$^{-3}$) molecular 
gas (\cite{Elit92}). The
line-of-sight velocity field also must be coherent (with respect to the sound
speed) on size scales $\gg 10^{13}$ cm to achieve significant amplification
(\cite{RM88}). The maser emission extends about $\pm300\kms$~from the galactic
systemic velocity of
1150\kms~adopted by Gallimore\etal (1996b). (Velocities throughout are
heliocentric and assume the optical astronomical definition of Doppler 
shift.) Gallimore\etal (1996b)
used the Very Large Array (VLA) of
the NRAO$^1$ to partially resolve the source structure on scales of 40 mas, 
or about one 

\noindent
\hrulefill

\noindent
The National Radio Astronomy Observatory is operated by Associated
Universities, Inc, under cooperative agreement with the National Science
Foundation.

\noindent
\hrulefill

\noindent
third of the
synthesized beamwidth, and inferred that the maser traces the midplane of 
the nuclear torus. Preliminary VLBI measurements also detected a
position-velocity gradient in the maser on angular scales of 0.4 mas
(\cite{Gw93}). We present the first VLBI synthesis images of the maser
emission at and redward of the systemic velocity.

\section{Observations and data}

We observed the H$_2$O maser emission in NGC~1068 for about 8 hours on 1994
November 5 with the Very Long Baseline Array (VLBA), and the VLA operating as
a phased 132 m aperture. The sensitivity of the VLA was critical to the
observations because of the low peak flux density of the maser, $\sim 0.6$ Jy.
We recorded bandpasses tuned to bandcenter heliocentric velocities of 1438.11,
1342.17, and 1165.80\kms. The velocity range 1210 to 
1285\kms~was not observed.  Figure~1 shows the velocity coverage of the
observations in the context of the maser spectrum. The channel spacing in each
band was 0.45\kms. We tracked variations in atmospheric phase using the maser
feature at 1462\kms as a reference and subtracted them from the data for 
other features.

Calibration and imaging relied upon standard spectral-line VLBI techniques.
We fit a 2-D Gaussian model to the brightness
distribution of every statistically significant peak in the images for each
velocity channel. The centroid positions, measured with respect to the
reference, are sensitive to calibration errors caused by
uncertainties in the station clocks and positions, earth orientation
parameters, and astrometric maser position. Only uncertainties in the last
two items are significant. These cause position errors that are linearly
proportional to the velocity offset from the reference. For the emission at
1325\kms~and 1135\kms, the errors are $\la 50~\mu$as and $\la 100~\mu$as,
respectively.

\section{Spectral-line images}

Five dominant clumps of maser emission are distributed almost linearly
on the sky, with pronounced velocity gradients (Figure~2). The emission close
to the systemic velocity (hereafter ``systemic emission'') is extended 
roughly east-west and lies to the southeast of the red-shifted emission. It
displays a velocity gradient of about 50\kms~mas$^{-1}$, with the higher
velocity emission in the west (Figure~3). The emission between about 1315 and
1465\kms~(hereafter ``red-shifted emission'') is distributed roughly in an arc
about 0.6 pc long, at a position angle of about $-45^\circ$. The red-shifted
emission is separable into clumps 
whose velocities
and velocity spreads increase with decreasing separation from the systemic 
emission. Clump 4, which lies closest to the systemic emission, 
has the greatest velocity spread ($\sim 100\kms$) but a reduced 
average velocity.
The uncertainties in measured positions 
after the data calibration are typically 100 times less than the characteristic
size scale of the maser source. The 0.4 mas gradient reported earlier 
by Gwinn\etal (1993), along a position angle of $75^\circ$ (note
correction of earlier typographical error), between 1391 and 1415\kms, is
consistent with the current observations.

No radio continuum emission was detected in our observations. An average of
all spectral channels redward of 1290\kms~shows no statistically significant
emission other than that which is related to the maser. The limiting flux
density is 0.45 mJy ($1\sigma$), corresponding to a brightness temperature of
$<5\times10^6$ K for a tapered beamwidth of $1.4 \times 1.0$ mas.
(We note that continuum emission coincident with clumps 1 -- 4 is only excluded 
at the 0.8 mJy level.)

\section{The molecular torus and the central engine}

We propose that the observed red-shifted maser emission traces part of the limb
of an edge-on rotating torus, rather than the midplane (cf. \cite{Galh2o}). This torus is
thick in both radial and axial directions and has an opening angle on the
order of $90^\circ$. Along the limb, the orbital motion is parallel to the
line of sight, and produces a substantial amplification along a
velocity-coherent path. Where the motion is transverse to the line of sight,
the limited thickness of the maser layer precludes significant amplification
except near the equatorial plane, where maser emission is perhaps visible perhaps because it
amplifies the 22 GHz continuum flux associated with the central engine or
radio jet. Observations that support this model are 1) the position angle of
the red-shifted emission relative to the radio jet axis; 2) a falling rotation
curve for cylindrical radii beyond $\sim 0.4$ pc; 3) the position-velocity
gradient of the systemic maser features; and 4) the location of red-shifted
maser emission, observed by us, with respect to the blue-shifted emission,
observed by Gallimore\etal (1996b). (The cylindrical radius is defined with
respect to an axis passing through the maser emission that lies at the
adopted systemic velocity, with a position angle of approximately $0^\circ$
(\cite{AHM94}; \cite{Galcont}).)

The model in Figure~4 shows the observed emission and the proposed location of
the blue-shifted emission. In the VLA map of the maser (\cite{Galh2o}) the
weighted average centroid of the blue-shifted emission between 800 and 
850\kms~lies 42 mas due east of the red-shifted emission at about 1425\kms, with
8 mas formal uncertainty in right ascension and declination. Based on this 
and the intrinsic symmetry of an edge-on torus, we hypothetically reflect the
red emission in velocity about the systemic velocity and in space about the
axis of the torus to obtain a ``V''-shaped structure.

A thick torus requires vertical support, possibly from internal velocities,
radiation forces, or magnetic fields. The ratio of internal motions to orbital
velocity, $v_i/v_\phi$, is approximately equal to the ratio of height to
radius, $h/R$ (\cite{KB88}).  The requisite velocity $v_i$ is on the order of
100\kms, corresponding to temperatures at which molecular gas could not
survive if the motions were thermal. However, the spread of velocities in
clumps 1 -- 4 approaches 100\kms. The source of the supersonic turbulence is
unclear, although our data show that internal motions increase dramatically
close to the central engine, suggesting that it could drive these motions.
Pier \& Krolik (1992) show that radiation forces acting on dust grains
close to the inner radius may be comparable to gravitational forces.
As well, gas densities of $10^{10}$
cm$^{-1}$ and magnetic field strengths of a few Gauss result in equipartition
of rotational and magnetic energy. Emmering, Blandford, \& Shlosman (1992) and
K\"onigl \& Kartje (1994) model centrifugally-driven flows from thin
magnetized accretion disks. At sufficiently large radii the flow may be dusty,
which is favorable to maser emission, though the model 
molecular gas density is too small.

The maser emission may arise from clouds that move in a warmer or
more turbulent medium.  Amplification occurs in gas with thermal motions of
less than a few \kms~but the turbulent velocities of the red-shifted emission are at
least an order of magnitude greater. The maser clouds may be the inward extension of
a system of molecular cloud cores that lie close to the
galactic plane at radii of at least 30 pc, with turbulent velocities of $\sim
100$\kms~(\cite{Tac94}; see also \cite{J93}), although the maser torus is misaligned with
respect to the galactic plane by approximately $65^\circ$.

The minimum radius at which maser emission is observed, $\sim 0.4$ pc, is
close to the radius at which dust sublimates (\cite{Bar87}; \cite{LD93}).
Graphite grains smaller than 0.05 $\mu$m sublimate within a radius of $r_{\rm
sub}\sim 0.4L_{45}^{0.5}$~pc, where $L_{45}$ is ultraviolet luminosity in
units of $10^{45}$ ergs s$^{-1}$ (\cite{Bar87}). Most of the intrinsic
bolometric luminosity is radiated at ultraviolet wavelengths. Hence, we adopt
$L_{45}=0.6$ (\cite{Pier94}), for which $r_{\rm sub}\sim0.4$ pc. Pier \& Krolik 
(1993) find the model infrared spectrum of a dusty torus best fits
the observed spectrum of NGC~1068 for a 0.5 pc inner radius and 800 K effective
temperature, somewhat less than the sublimation temperature of silicate dust. The
agreement of $r_{\rm sub}$ with the maser observations may be related to the
important role that dust can play in the thermodynamics of the maser pump
cycle (e.g., \cite{CW95}) and in obscuring the nuclear continuum. 
Ablation of the component clouds of the torus
by photons from the central engine (\cite{KB86}, 1988; \cite{PV95}) also fixes
the inner radius to be on the order of 1 pc. 

The radial extent of maser emission is probably governed by limitations in the
maser excitation mechanism operating in the torus clouds. Emission can be
driven by wind or jet-induced shocks (\cite{EHM89}, \cite{KN96}), or by
collisional heating associated with X-ray illumination by the central engine,
as proposed for NGC~4258 and NGC~1068 (\cite{NMC94}). Both mechanisms excite
only a thin layer of material. As well, curvature of the torus may block
direct mechanical interaction or X-ray illumination at large radii. 

The maser
torus model suggests that there should be red-shifted emission visible from the
southern half of the torus and systemic emission superposed along some 
length of the jet.
The absence of southern emission may be related to the absence of strong
southern jet emission, on scales smaller than about 10 pc (e.g., \cite{UNW87}).
Alternatively, southern radio emission may be preferentially obscured,
especially if the torus is not strictly edge-on. The presence of systemic
maser emission near the equator alone is consistent with the jet having a
steep spectrum and a rapid decline with radius of the frequency at which it
becomes optically thin. In NGC~4258, the 22 GHz jet emission is localized to
radii on the order 0.2 mas scaled to a distance of 15 Mpc
(\cite{Herrn96}). Since background continuum emission has not been
observed directly by us, the relatively flat-spectrum nuclear source seen on
angular scales of $\sim 0\rlap{.}''1$ (\cite{Galcont}; \cite{Mux96}) cannot be
entirely a self-absorbed synchrotron source.  The emission is not limited to a
point source; it could represent thermal emission or possibly Thompson-scattered
radiation (\cite{Galcont}). Background amplification may still give rise to 
the systemic maser features because they are relatively weak. The limit on 
continuum strength implies the gain is at least 40, which is modest.

>From the rotation curve in Figure 3, we estimate the total mass enclosed
within a radius of 0.65 pc is $\sim 1\times10^7$ M$_\odot$, assuming a
circular orbital velocity of $\sim 250$\kms. The luminosity of the central
engine (\cite{Pier94}) is $\sim 0.5$ times the corresponding Eddington
luminosity. The mass is less than the approximately $2\times10^8$ M$_\odot$ implied by
the motions of the molecular cloud cores beyond a radius of 30 pc
(\cite{Tac94}). From VLA observations of the maser Gallimore\etal (1996b)
estimate a mass of $3\times10^7$ M$_\odot$ (scaled to 15 Mpc), though the VLBA
observations provide a more accurate determination of the rotation curve. The
VLBA mass estimate is uncertain by factors of order unity because the best-fit
rotation curve is sub-Keplerian, $v\propto R^{-0.31\pm0.02}$ for $0.65>R>0.40$
pc. Self-gravity of the torus or substantial nearby stellar mass could be
responsible. Alternatively, outwardly directed radiation pressure may reduce
orbital velocities close to the inner edge of the torus (\cite{PK92}). This
proposed force and the known dispersion in maser velocities is smallest for 
0.65 pc radius, at which we estimate the central mass.

If the {\it systemic} masers are restricted to the inner surface of the torus,
then the systemic emission arises over a narrow range of radii. The changing
line-of-sight projection of the orbital velocity causes a gradient in
line-of-sight velocity (as a function of impact parameter in the equatorial
plane), ${\partial V / \partial b} = 0.3 M_6^{0.5}r_{\rm pc}^{-1.5}D_{\rm Mpc}
\kms~{\rm mas}^{-1}, $ where $M_6$ is the central mass in units of $10^6$
M$_\odot$, $r_{\rm pc}$ is the radius in parsecs, and $D_{\rm Mpc}$ is the
distance in Mpc. For parameters estimated previously from the red-shifted
emission, $M_6=10$, $r_{\rm pc}=0.4$, and $D_{\rm Mpc}=15$, we predict a
gradient of $60\kms$~mas$^{-1}$, in reasonable agreement with observation
(Figure~3).  Greenhill\etal (1995b) apply similar arguments to the maser disk
in NGC~4258. This interpretation of the systemic emission would be bolstered
if the line-of-sight velocities of individual systemic maser features are
observed to drift in time, corresponding to the line-of-sight centripetal
acceleration 
while the red and blue-shifted maser features remain fixed 
(cf., \cite{Gre95drift}, \cite{Nak95} for NGC~4258).  The large
velocity dispersion of the innermost maser clump (no. 4) may in part reflect a
turnover in the rotation curve, rather than just high internal motions. The
turnover would correspond to the reduced line-of-sight projections of the orbital
velocity of masers that lie near the inner edge of the torus and out of the
sky plane that contains the red-shifted emission.

\section{Summary}

We have mapped the brightness distribution of H$_2$O maser emission at and
redward of the systemic velocity in the Seyfert nucleus of NGC~1068, with
sub-milliarcsecond resolution. The linear
structure of the maser is misaligned with respect to the axes of the radio
synchrotron jet and associated optical ionization cone. The observed maser
emission can be reflected about the systemic velocity and jet axis to yield a
``V''-shaped structure on the sky, and this may be interpreted as tracing the
limb of a thick torus. Observations of the blue-shifted maser emission,
with lower angular resolution, are consistent with this interpretation.
Velocity dispersions up to $\sim
100$\kms~may indicate that the torus probably consists of individual maser clouds
immersed in a warmer more turbulent medium.

The enclosed mass within a radius of 0.65 pc is $\sim 1\times10^7$ M$_\odot$,
and the rotation curve is sub-Keplerian. The estimated mass is consistent with
a position-velocity gradient in the systemic maser emission if the systemic
emission arises on the inner edge of the torus at a radius of 0.4 pc. For this
mass, the central engine of NGC~1068 radiates about 0.5 times its Eddington
luminosity. No radio continuum source was detected toward the center of the torus,
which argues that the flat-spectrum continuum emission observed on larger scales 
is not the byproduct of a self-absorbed synchrotron source.

\acknowledgments

We are grateful to O. Blaes, J. Gallimore, and J. Krolik for helpful comments
and discussions. M. Eubanks and J. R. Herrnstein provided antenna positions.
We thank P. Diamond and J. Benson for assistance with an early version of the
correlator model. This work was supported in part by the National Science
Foundation (AST92-17784).

\clearpage

\begin{figure}

\plotfiddle{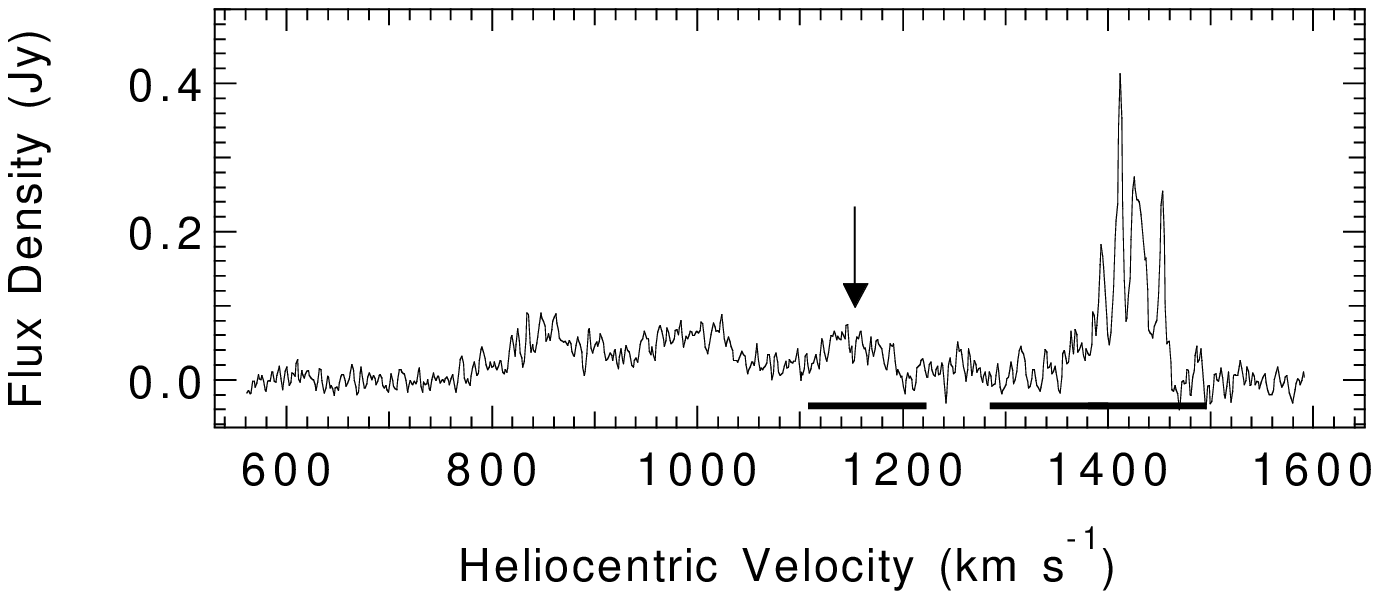}{3.0in}{0}{100}{100}{-200}{-600}
\caption{Spectrum of the H$_2$O maser in the nucleus of NGC~1068, taken with the
Effelsberg 100-m antenna in 1993 February. The flux density calibration is
accurate to about 30\%. The arrow indicates the systemic velocity of the 
galaxy.  Solid bars indicate the ranges of velocities we observed.}

\end{figure}

\begin{figure}

\plotfiddle{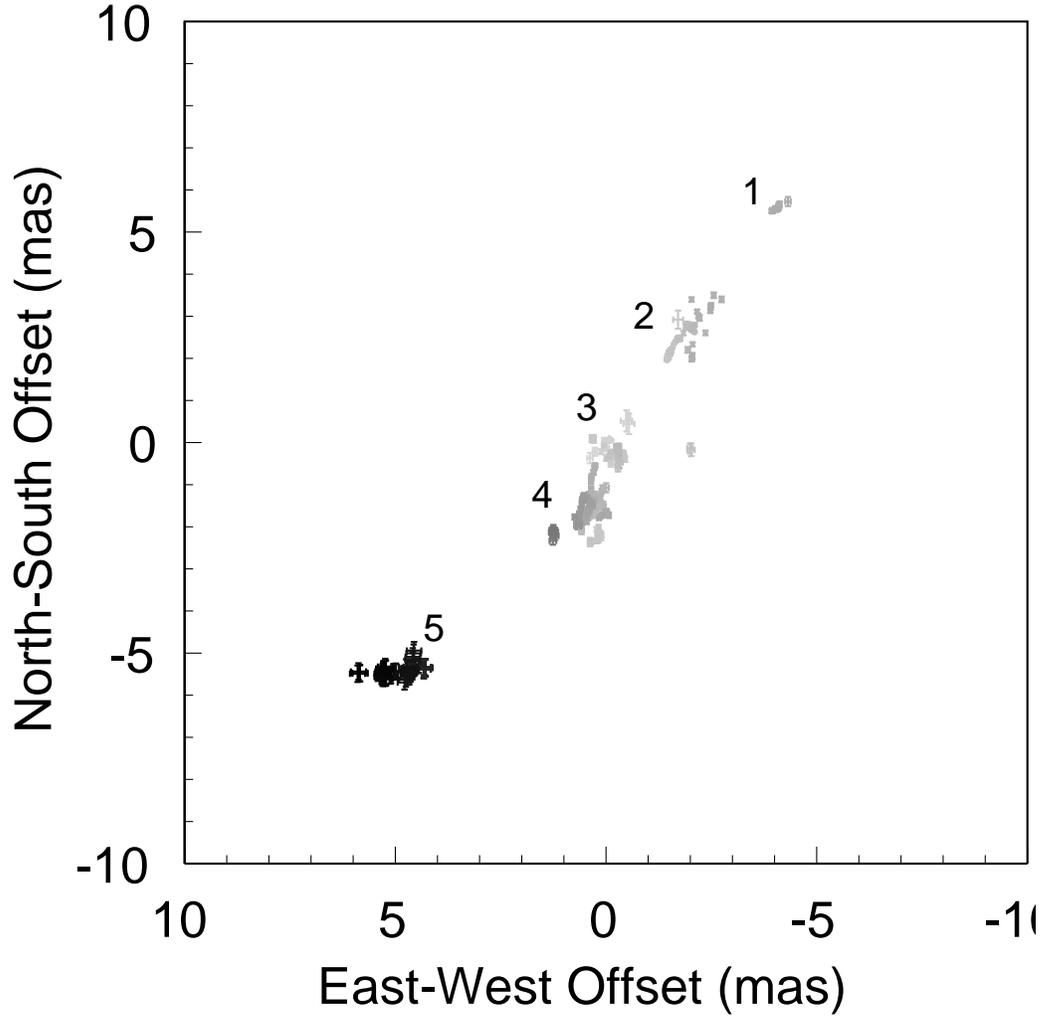}{5.0in}{0}{80}{80}{-315}{-130}
\caption{Map of the brightness distribution of the maser emission. The positions
are relative to the emission at 1462\kms~and velocities are indicated by
gray scale such that more red-shifted velocities are lighter. Error bars include both random and
systematic measurement errors for each emission component in each spectral
channel. The southeastern complex (no. 5), elongated east-west, is the emission within
about 10\kms~of the systemic velocity. The rest of the emission is distributed
along a
position angle of about $-45^\circ$ and is red-shifted by 150--300\kms. 
The blue-shifted emission between about 800
and 1100\kms~was not observed.}

\end{figure}

\begin{figure}

\plotfiddle{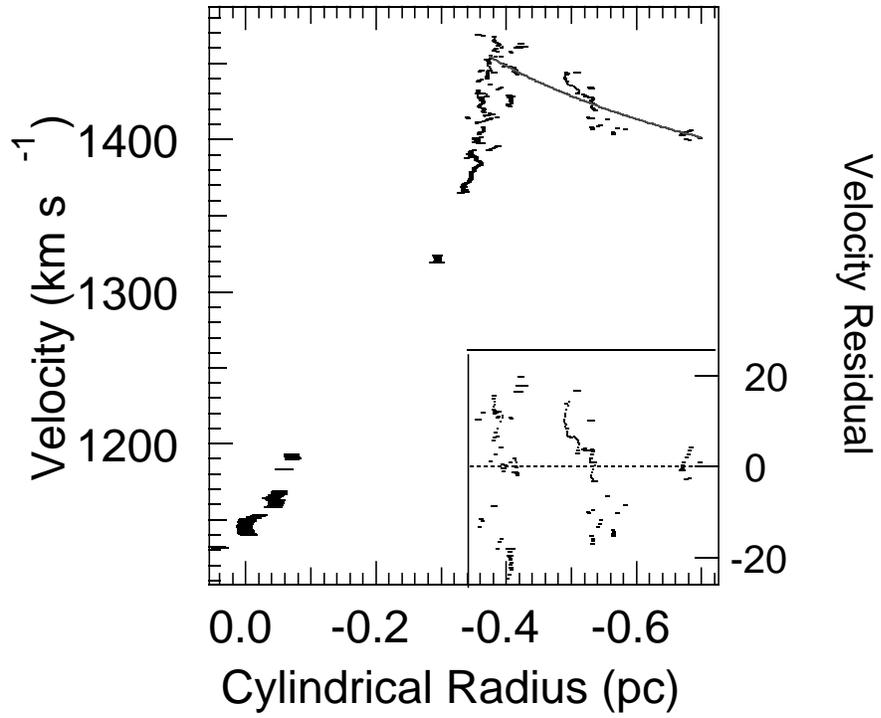}{4.5in}{0}{120}{120}{-180}{-650}
\caption{Position-velocity diagram of the maser emission. The cylindrical
radius is measured with respect to the maser feature at the systemic velocity along
a position angle of $90^\circ$.
Error bars indicate both random and systematic measurement errors. The dashed
rotation curve is discussed in section 4. 
({\it insert}) Residuals from the fitted curve.}

\end{figure}

\begin{figure}

\plotfiddle{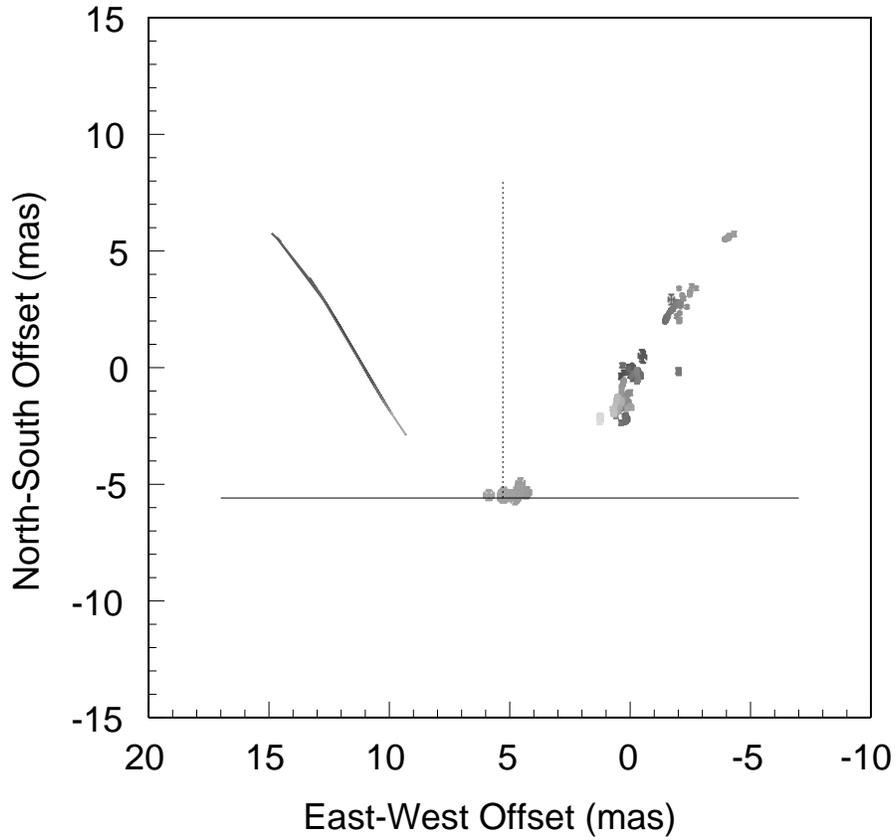}{5.0in}{0}{80}{80}{-300}{-155}
\caption{Model showing the location of the observed maser emission, as in
Figure~2, and the anticipated location of the emission blue-shifted with
respect to the systemic velocity. Color indicates line-of-sight velocity. The
vertical line indicates the axis of the radio jet.  The maser emission traces
the limb of the upper half of an axially thick torus for radii between about
0.4 and 0.65 pc.}

\end{figure}

\end{document}